\documentclass{iau}
\usepackage{graphicx}
\usepackage{color}
\usepackage{natbib}
\newcommand{\aap}{    {\it Astron. Astrophys.}}
\newcommand{\apj}{    {\it Astrophys. J.}}
\newcommand{\apjl}{    {\it Astrophys. J. Lett.}}

\newcommand{\mnras}{   {\it Mon. Not. Roy. Astr. Soc.}}

\newcommand{\solphys}{{\it Solar Phys.}}

\newcommand{\pasj}{\it Publ.~Astr.~Soc.~Japan}

\title[Rogue Active Regions and the Inherent Unpredictability of the Solar Dynamo] %% give here short title %%
{Rogue Active Regions and the Inherent Unpredictability of the Solar Dynamo}

\author[K. Petrovay \& M. Nagy]   %% give here short author list %%
{Krist\'of Petrovay \and Melinda Nagy}

\affiliation{Department of Astronomy, E\"{o}tv\"{o}s Lor\'and University, 
Budapest, Hungary \\ email: {\tt K.Petrovay@astro.elte.hu, M.Nagy@astro.elte.hu} }

\pubyear{2018}
\volume{340}  %% insert here IAU Symposium No.
\jname{Long-Term Datasets for the Understanding of Solar and Stellar Magnetic Cycles}
\editors{Dipankar Banerjee, Jie Jiang, Kanya Kusano, Sami Solanki, eds.}

\begin{document}

\maketitle

\begin{abstract}
New developments in surface flux transport modeling and theory of flux
transport dynamos have given rise to the notion that certain large
active regions with anomalous properties (location, tilt angle and/or
Hale/non-Hale character) may have a major impact on the course of
solar activity in subsequent years, impacting also on the amplitude of
the following solar cycles. Here we discuss our current understanding
of the role of such ``rogue'' active regions in cycle-to-cycle
variations of solar activity.
\end{abstract}

\firstsection % if your document starts with a section,
\section{Dynamo-based cycle forecasting and rogue active regions}

In the last ten years significant advance has been made towards solar
cycle forecasting based on dynamo models. Most efforts in this
direction are based on the flux transport dynamo concept. An essential
feature of these models is that the poloidal magnetic field, peaking
near the poles around the minimum of the solar cycle, serves as seed
for the toroidal field built up in the next cycle: the strength of the
polar field is therefore a good predictor of the amplitude of the next
solar cycle. This is confirmed by a good empirical correlation between
the respective indicators (e.g. \citealt{Munoz2013PRL}). The problem
of predicting an upcoming solar cycle is then reduced to the problem
of predicting the peak strength of the polar field being built up
during the course of the ongoing cycle. Surface flux transport models
based on observations indicate that the polar magnetic flux is built
up from the {\it unbalanced} trailing polarity flux originating in
active regions of the ongoing cycle. As the leading and trailing
fluxes in a bipolar magnetic region are initially balanced, a
significant contribution to the polar flux is only expected if the two
polarities are located at significantly different latitudes,
facilitating a more effective cancellation of flux of one (typically,
leading) polarity with its opposite hemisphere counterpart across the
equator. This is, in turn, easier to achieve at low latitudes and for
relatively high tilts.

A single bipolar magnetic region represents a contribution
\begin{equation}
 \delta D_{\mathrm{BMR}} \approx F \,d \, \sin\alpha \,  \sin\theta
 \label{eq:thenumber}
\end{equation}
to the solar axial dipole moment where $F$ is magnetic flux, $d$ is
the separation of the two polarities, $\alpha$ is the tilt angle,
$\theta$ is the colatitude. Active regions with unusually high or
deviant values of the parameters may then be expected to induce
significant fluctuations in the strength of the polar magnetic field
built up in a cycle. In addition to {\it tilt quenching,} i.e. a
nonlinear dependence of the mean value of the tilt angle $\alpha$ on
cycle amplitude (\citealt{Dasi2010}), a  possibly important factor in
intercycle variations are fluctuations in the unbalanced flux
contribution by active regions, related to the random nature of the
flux emergence process.  Indeed, while \cite{Cameron2010} find that
tilt quenching satisactorily reproduces the observed polar field
amplitude built up during cycles 15 to 22, the same approach fails for
Cycle 23.  \cite{Cameron2013} suggested that the vagaries of the flux
emergence process can be responsible for such deviations from the
expected behaviour of the solar cycle. This was corroborated by
\citeauthor{Jiang2014} (\citeyear{Jiang2014}, \citeyear{Jiang2015})
who showed that assimilating individual active regions into the model,
the polar field amplitude built up in Cycle 23 can be reproduced.

This suggests that in extreme cases even a single large active region
with unusual properties can have a dramatic impact on the further
course of cyclic solar magnetic activity. To have such a major impact,
an active region needs to be (1) large, (2) unusually tilted, strongly
deviating from Joy's law (e.g. non-Joy or very ``over-Joy'') (3)
positioned at low latitudes to facilitate the cross-equatorial
cancellation of flux of one polarity. In a recent work
(\citealt{Nagy2017}) we reported examples of potentially dramatic
effects of such  ``rogue'' active regions on the solar cycle in a
dynamo model. The main results from this research are summarized in
the next section, while the third, concluding section briefly
discusses the implications of these results for the importance of
long-term datasets.

\section{Rogue active regions in the $2\times2$D dynamo}

Dynamo models incorporating individual active regions have been
developed by several research groups (\citealt{Yeates2013},
\citealt{Yeates2015}, \citealt{Miesch2016}, \citealt{Hazra2017}). One
particularly attractive approach is the $2\times2$D dynamo
developed in Montreal (\citealt{Lemerle2015}, \citealt{Lemerle2017}).

The model couples a 2D surface flux transport simulation (SFT)  with a
2D axisymmetric flux transport dynamo (FTD).  The azimuthally averaged
SFT component provides the upper boundary condition for the FTD
component, while the FTD module couples toward the SFT by the
emergences of new bipolar magnetic regions (BMR). This step is based
on a semi-empirical emergence function that gives the emergence
probability of a BMR depending on the toroidal magnetic field $B_\phi$
at the bottom of the convective zone in the FTD module. In the
optimized solution used here the emergence probability is proportional
to $B_\phi^{3/2}$; there is, however, a  threshold below which flux
emergence is suppressed. Properties of the emerging BMR --- flux,
angular separation, tilt --- are randomly drawn from distribution
functions for these quantities built from observed statistics of solar
active region emergences (see Appendix A in \citealt{Lemerle2015}). 

The other nonlinearity built in the model is tilt quenching: a
reduction of the average BMR tilt angle $\alpha$ with $B_\phi$
according to the ad hoc formula
\begin{equation}\label{eq:tiltquenching}
    \alpha = \frac{\alpha_0}{1 + (B_\phi/B_q)^2},
\end{equation}
where $B_q$ is the quenching field amplitude.

The main advantages of this model are its high numerical efficiency
and the fact that it is calibrated to follow accurately the
statistical properties of the real Sun. The complete
latitude--longitude representation of the simulated solar surface in
the SFT component further makes it possible to achieve high spatial
resolution and account for the effect of individual active region
emergences. 

The reference solar cycle solution presented in \citet{Lemerle2017},
which is adopted in the numerical experiments of \citet{Nagy2017} is
defined by 11 adjustable parameters, which were optimized using a
genetic algorithm designed to minimize the differences between the
spatiotemporal distribution of emergences produced by the model, and
the observed sunspot butterfly diagram. Thanks to the numerical
efficiency of the model, the reference solution can be run for many
hundreds of cycles, in contrast to the limited number of actual solar
cycles observed. We have studied this reference solution looking for
sudden changes in the behaviour of the dynamo and trying to identify
the culprits.  We found that in some cases even a single rogue BMR can
have a major effect on the further development of solar activity
cycles, boosting or suppressing the amplitude of subsequent cycles. In
extreme cases an individual BMR can completely halt the dynamo,
causing a grand minimum. Alternatively a dynamo on the verge of being
halted can also be resuscitated by a rogue BMR with favourable
characteristics. Rogue BMR also have the potential to induce
significant hemispheric asymmetries in the solar cycle. 

\begin{figure}%[t!]
  \centering
  \includegraphics[width=\linewidth]{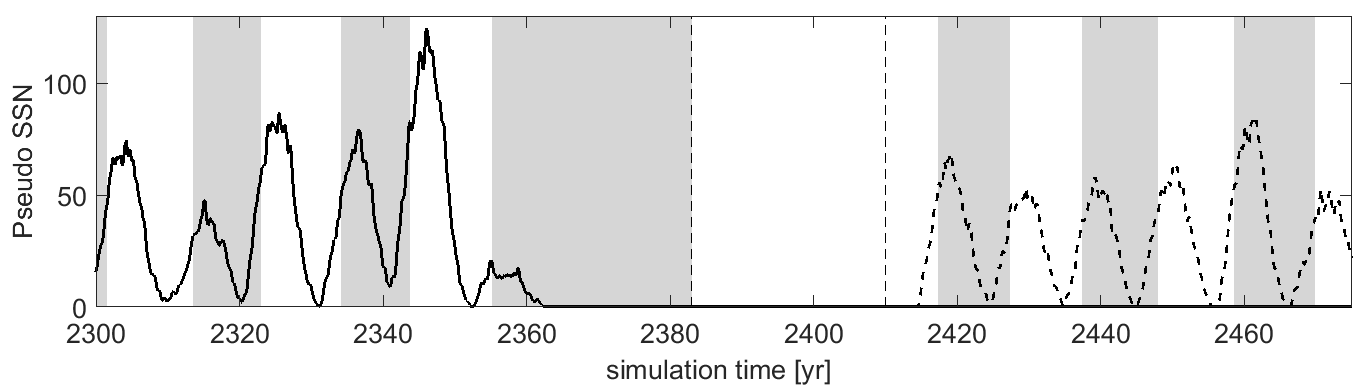}\\
  \includegraphics[width=\linewidth]{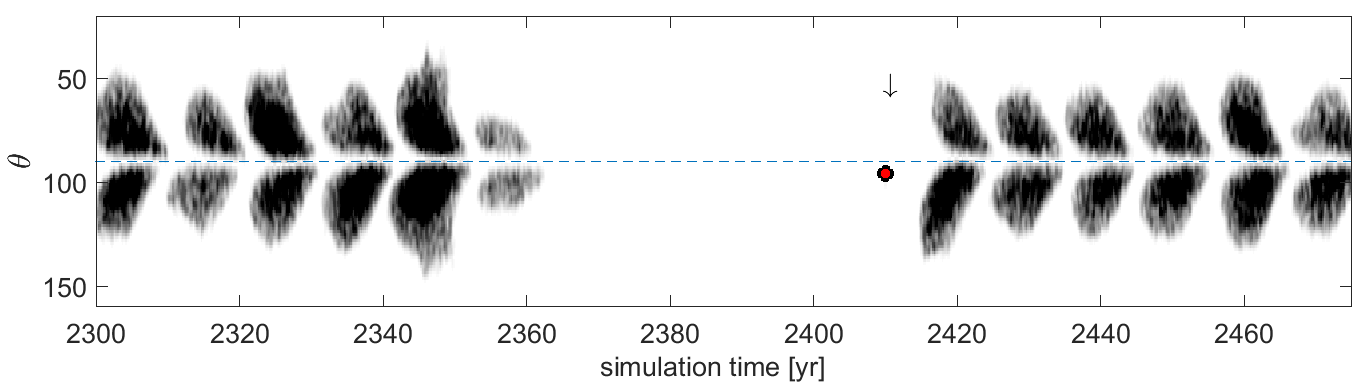}\\
  \caption{\label{fig:grandmin} A rogue sunspot restarting cyclic
  activity  
  after a 60 years long grand minimum state. 
  On the \emph{top panel} black dashed line shows the cycles launched
by a single BMR manually inserted on the souther hemisphere, close to
the equator. The gray background corresponds to the negative dipole
periods, while the vertical dashed lines show the 30 years long
period, when it was zero.  On the \emph{bottom panel} we show the
butterfly diagram of the simulation run with the BMR inserted. The
position of the leading polarity is indicated by the red dot. The
properties of this BMR are listed in the first row of Table
\ref{tab:BMRs}.}
\end{figure}

In addition to these effects, discussed  in detail in \cite{Nagy2017},
here we present a further possible role of rogue active regions.
While, owing to the presence of the field strength threshold, the
dynamo is not self-excited and it can never recover from a grand
minimum state, in one case, displayed in  Figure \ref{fig:grandmin},
we manually insert a large rogue BMR in the model at $t = 2410$ during
a grand minimum phase, overriding the threshold. A strong ``over-Joy''
BMR with similar properties (given in the first data row of Table 1)
did indeed arise in one of the model runs (see Figure 4 of
\citealt{Nagy2017}). As apparent from the figure, this BMR is capable of
inducing a recovery even from a long lasting deep grand minimum state.

\begin{figure}
  \centering
  \begin{minipage}{0.55\textwidth}
  \includegraphics[width=0.95\linewidth]{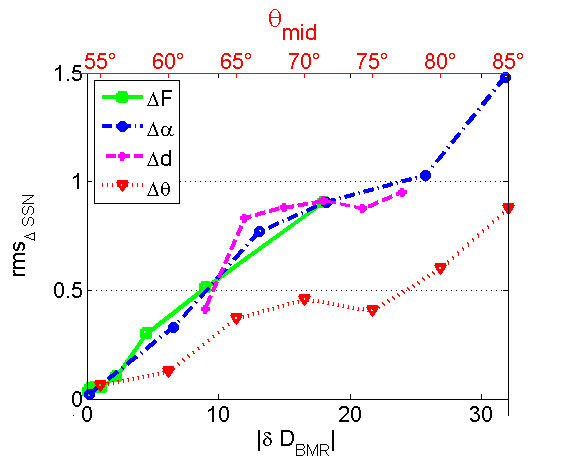}
  \end{minipage}
  \begin{minipage}{0.44\textwidth}
  \caption{Average effect of varying the properties of a BMR (2nd data
row of Table \ref{tab:BMRs}), inserted in the simulations at cycle
maximum, on the amplitude of the subsequent cycle. Variations in BMR
flux (green), tilt angle (blue) and  polarity separation (magenta) are
converted to the contribution to the dipole moment (in units of
$10^{21}\,$Mx) according to
Equation \ref{eq:thenumber}, while the varying colatitudes (red) are
shown on the top axis.}\label{fig:summarize}
  \end{minipage}
\end{figure}

Figure \ref{fig:summarize} summarizes the results of several numerical
experiments that aimed to study how the characteristics of active
regions affect the subsequent, or even the ongoing cycle
\citep{Nagy2017}. For this analysis a 'test' BMR was selected with
characteristics listed in the second data row of Table \ref{tab:BMRs}.
(Such a BMR did also emerge spontaneously during the reference
simulation run.) This active region was manually inserted into ongoing
simulations with preset parameters. The experiments were performed for
three cycles of average, below average and above average,
respectively. In each case two series of experiments were carried out
with Hale (anti-Hale) test-BMR in order to increase (decrease) the
dipole moment of the examined cycle. The characteristics of the
test-BMR -- emergence time and latitude, flux, tilt angle and angular
separation -- were changed one by one in order to map the impact of
each property on the subsequent simulated cycle.

As it is shown in Figure \ref{fig:summarize}, the flux, tilt angle and
separation (green, blue and magenta curves) have quite similar effect
on the temporal evolution of the dipole moment, and consequently of
the next cycle. The good agreement of the green, blue and magenta
curves in the plot confirms that the effect of these factors can
indeed be combined in the form given in equation
(\ref{eq:thenumber}) providing a good measure of the ``dynamo
efficiency'' of individual active regions at a given latitude.

The red curve, showing the influence of the emergence latitude,
indicates that the effect of a BMR decreases with the emergence
latitude. Nevertheless, BMRs appearing $20^{\circ}$ far from the
equator can still have significant impact on the next cycle in the
present model (but see discussion in the following section).

By setting the emergence time \citet{Nagy2017} found that active
regions emerging during the rising phase of the cycle can modify even
the ongoing cycle, but this effect disappears at cycle maximum. The
strongest impact on the subsequent cycle is obtained if the test BMR
emerges at cycle maximum while for later epochs the effect gradually
subsides.

%\begin{figure}[t!]
%  \centering
%  \includegraphics[width=\linewidth]{experiment_35_addF139_tocycle09_positive_DBeta.png}\\
%  \caption{\label{fig:Dbeta}Tilt angle of an inserted rogue BMR and its effect on the amplitude of the subsequent cycle, similarly to Figure 9 in \citet{Nagy2017}. Gray background indicate intervals of negative dipole moment. The properties of this BMR are listed in the second row of Table \ref{tab:BMRs}.}
%\end{figure}

\begin{table}[h!]
    \center
  \begin{tabular}{| c | c | c | c | c | c | c | c |}
  \hline
  $\theta_{\mathrm{lead}}$ & $\theta_{\mathrm{trail}}$ & $F $ [$10^{23}$ Mx] & $\alpha$  & $d$ & $\delta D_{\mathrm{BMR}}$[$10^{23}$ Mx] & J/H & \\
  \hline
  95.6$^{\circ}$           & 104.1$^{\circ}$           & --3.58              & $-15.53^{\circ}$ & 32.11$^{\circ}$   & 0.5293 &  J/H & Figure \ref{fig:grandmin}\\
  \hline
  89.5$^{\circ}$           & 82.1$^{\circ}$            & --1.39              & 13.98$^{\circ}$  & 30.97$^{\circ}$   & --0.1810 & J/H, J/a-H & Figure \ref{fig:summarize}\\
  \hline
  \end{tabular}
%	$\theta_{\mathrm{lead}}$           & 95.6$^{\circ}$	   & 89.5$^{\circ}$    \\
%	$\theta_{\mathrm{trail}}$          & 104.1$^{\circ}$   & 82.1$^{\circ}$     \\
%	$F $ [$10^{23}$ Mx]                & --3.58            & --1.39	         \\
%	$\alpha$                           & $-15.53^{\circ}$  & 13.98$^{\circ}$    \\
%	$d$                                & 32.11$^{\circ}$	& 30.97$^{\circ}$   \\
%    $\delta D_{\mathrm{BMR}}$[$10^{23}$ Mx] & 0.5293       & --0.1810 \\
%    J/H                                & J/H              & J/H, J/a-H\\
%    \hline
%	                          & Figure \ref{fig:grandmin} & Figure \ref{fig:Dbeta}
%  \end{tabular}
  \caption{Parameters of active regions discussed in the
paper.  Colatitudes $\theta_{\mathrm{lead}}$ and
$\theta_{\mathrm{trail}}$ are the
  latitudinal positions of leading and trailing polarities;  $F$ is
the flux of the trailing polarity ($F_{\mathrm{trail}} =
-F_{\mathrm{lead}}$); $\alpha$ is the tilt angle and $d$ is the
angular separation of leading and trailing polarities. $\delta
D_{\mathrm{BMR}}$, the contribution of the BMR to the global dipole
moment, is defined according to Equation \ref{eq:thenumber}. J/H
indicates whether the active region is (anti-)Joy/(anti-)Hale. In the
case of the second row a J/H (J/a-H) test-BMR increases (decreases)
the dipole moment during the experiments detailed in Section 5 of
\citep{Nagy2017}. }\label{tab:BMRs}
\end{table}

\section{Implications for observational records and their analysis}

It is to be noted that the magnetic fluxes of the rogue BMR identified
in the simulations are mostly within or only slightly outside the size
range of solar active regions on record. While for historical records
the magnetic fluxes can only be roughly estimated based on sunspot
areas, such estimates suggest that the largest spots on record were in
the $(1$--$3.5)\cdot 10^{23}\,$Mx range. What seems to be more
exaggerated is the polarity separations: in fact, observed values of
$d$ do not normally exceed about $20^\circ$. The high values in the
simulations are due to the fact that \cite{Lemerle2015} calibrate
their $d$--$F$ relation to cycle 21 which did not show any AR with
flux exceeding $10^{23}\,$Mx, so the application to rogue spots is
based on a somewhat dubious extrapolation. Nevertheless, equation
(\ref{eq:thenumber}) indicates that a 33\,\% reduction in $d$ is
readily compensated by, e.g. a 50\,\% increase in $F$ which is still
not inconceivable. So while the occurrence rate of rogue AR might be
overestimated in the model, the reality of the phenomenon is not
brought into question.

In the CESAR Solar Data Archives maintained by Kanzelh\"ohe
Observatory  a list of the largest sunspots recorded in the
observatory is 
presented.\footnote{\tt
http://cesar.kso.ac.at/spots/biggest\_spots.php \hfill\strut} 
It is remarkable that 5 out of the 6 largest
spots on the list arose in the period 1946--1951, i.e. during the
declining phase of Cycle 18. It is well known that Cycle 19 that
followed this was by far the strongest and most anomalous cycle on
record; reproducing its amplitude represents a challenge for dynamo
based cycle prediction methods. 

A detailed analysis of the effect of these and similar large active
regions on the dynamo is, however, still hindered by a number of
unresolved issues both from the theoretical and observational side. 

On the theoretical side, equation (\ref{eq:thenumber}) only gives the
{\it initial} contribution of a BMR to the global dipole. The final
net contribution will be determined by the amount of unbalanced flux,
i.e. by the amount of flux cancelling across the equator during the
subsequent evolution of the BMR. The relation between initial and
final contributions has been studied by \cite{Jiang2014}; however,
this relation may be quite sensitive to model details, possibly
explaining why the latitude dependence reported by \cite{Nagy2017}
differs. This dependence on model details is currently under study; a
more reliable measure of the latitude dependence of the dynamo
efficiency of active regions can only be constructed after the
clarification of this issue.

From the observational point of view, detailed and reliable
information on the evolution of large active regions (rogue active
region candidates) is needed. In currently available historical
records even the reported sunspot areas often differ by a large amount
between various sources. Magnetic flux and even magnetic polarity data
are not always well correlated with the white light information on
sunspot groups and they are only sporadically available for epochs
prior to the invention of magnetographs. Collecting and correlating
all relevant information from all available sources is therefore
important for a reliable assessment of the dynamo effectivity of large
active regions and for identifying rogue active region candidates.

Finally, the largest active regions often live for several Carrington
rotations, displaying dramatic changes in extent, appearance and spot
distribution. During this extended period of time the constituent
sunspots only gradually decay to release weak magnetic flux into the 
plage where the linear transport processes modelled in SFT simulations
will take over. Representing active regions as a simple bipolar pair
of instantaneous point sources is clearly a very crude approximation
to this process. More realistic models of active region sources of
magnetic flux in SFT simulations need to be developed, and for these
every piece of information concerning the detailed structure and
evolution of large active regions can potentially have a high
significance.

As a final remark we note that unusually large active regions
harbouring a significant amount of free energy may also be responsible
for superflares, as recently discussed by \cite{Shibata2013},
\cite{Maehara2017} or \cite{Namekata2017}. As tilt and free energy are
closely related to the writhe and twist of magnetic flux bundles,
respectively, which are but two manifestations of magnetic helicity,
the categories of rogue AR and superflaring AR may overlap to a
significant extent. Yet we stress that size alone does not imply that
an active region will have a major influence: it is the added presence
of helicity, manifest as tilt and/or free energy, that can lead to a
major effect of sunspots on either space weather or space climate.

\section*{Acknowledgements} 
K.P. thanks the symposium organizers for the invitation and their
hospitality during the meeting. Our research on the solar dynamo is
supported by the Hungarian National Science Research Fund (OTKA grant
no.\ K128384). M.N.'s research is currently funded by the 
\'UNKP-16-3 New National Excellence Program of the Ministry of Human
Capacities of Hungary. The paper made reference to sunspot data
provided by Kanzelh\"ohe Observatory, University of Graz, Austria.

%\bibliographystyle{astron}
%\bibliography{References_IAUS340_PKNM}

\end{document}